\documentclass[aps,prd,preprint,superscriptaddress,showpacs,nofootinbib]{revtex4-1}
\usepackage{latexsym}
\usepackage{amssymb, amsmath}
\usepackage{epsfig, graphics,pstricks, color}
\numberwithin{equation}{section}

\newcommand{\beq}{\begin{equation} } 
\newcommand{\eeq}{\end{equation} } 
\DeclareMathOperator{\Tr}{Tr}

\begin{document}

\baselineskip=21pt \pagestyle{plain} \setcounter{page}{1}

\vspace*{-1.1cm}

\begin{flushright}{\small Fermilab-PUB-13-549-T}\end{flushright}

\vspace*{0.2cm}

\begin{center}

{\Large \bf  $W'$ signatures with odd Higgs particles}\\ [9mm]

{\normalsize \bf Bogdan A. Dobrescu$^\star$ and Andrea D. Peterson$^{\star\diamond}$  \\ [4mm]
{\it
$\star$ Theoretical Physics Department, Fermilab, Batavia, IL 60510, USA \\ [2mm]
$\diamond$ Department of Physics, University of Wisconsin, Madison, WI 53706, USA
}\\
}

\vspace*{0.5cm}

February 16, 2014

\vspace*{0.7cm}

{\bf \small Abstract}

\vspace*{-0.2cm}

\end{center}

We point out that $W'$ bosons may decay predominantly into Higgs particles 
associated with their broken gauge symmetry. We demonstrate this in a renormalizable model
where the $W'$ and $W$ couplings to fermions differ only by an overall normalization. 
This ``meta-sequential" $W'$ boson decays into a scalar pair, with the charged one 
subsequently decaying into a $W$ boson and a neutral scalar. 
These scalars are odd under a parity of the Higgs sector, which consists of a complex bidoublet and a doublet.
The $W'$ and $Z'$  bosons have the same mass and  branching fractions into scalars, and may show up at the LHC in  final states involving one or two electroweak bosons and missing transverse energy. 
\\ [-1mm]

\nopagebreak
\section{Introduction}

New heavy particles of charge $\pm 1$ and spin 1, referred to as $W'$ bosons, are predicted in many interesting theories for 
physics beyond the Standard Model (SM) \cite{Beringer:1900zz,Mohapatra:1986uf}. 
Extensive searches for $W^\prime$ bosons at 
colliders have set limits on the production cross section times branching fraction in several final states \cite{Beringer:1900zz}.
The most stringent limit on a $W'$ boson that has the same couplings to quarks and leptons as the SM $W$ boson
(``sequential" $W'$) has been set using the $\ell\nu$ channels, where $\ell = e$ or $\mu$; the current mass limit 
is 3.8 TeV, set by the CMS Collaboration \cite{CMS:2013rca} using the full data set from the 8 TeV LHC. 

In this paper we show that the $W'$ boson is likely to decay not only into SM fermions, as often assumed,
but also into pairs of scalar particles from the extended Higgs sector responsible for the $W'$ mass. 
As a result the existing limits may be relaxed, and different types of
searches at the LHC may prove to be more sensitive.

Theories that include a $W'$ boson embed the electroweak gauge group 
within an  $SU(2)_1\times SU(2)_2\times U(1)$, 
$SU(3)_W\times U(1)$, or larger
gauge symmetry that is spontaneously broken down to the electromagnetic gauge group, $U(1)_{\rm em}$.
This symmetry breaking pattern is induced usually by some scalar fields with
vacuum expectation values (VEVs). The coupling of the $W'$ to these scalars 
is related to the gauge couplings, and cannot be too small.
In perturbative renormalizable models,
the scalars have masses near or below the symmetry breaking scale, because the quartic 
couplings grow with the energy. The $W'$ boson, by contrast, may be significantly heavier,
because large gauge couplings are allowed by the asymptotic freedom of 
non-Abelian gauge theories. Consequently, it is natural to expect $W'$ decays into pairs of
particles from the extended Higgs sector.

We demonstrate the importance of $W'$ decays into scalars by analyzing in detail a simple
renormalizable $W'$ model: $SU(2)_1\times SU(2)_2\times U(1)_Y$ gauge symmetry broken by the VEVs
of two complex scalars: a bidoublet ({\it i.e.}, a doublet under each non-Abelian group) of hypercharge $Y = 0$
and a doublet under one of the $SU(2)$'s. 
This model has been studied in different contexts \cite{Barger:1980ix,Cao:2012ng}, assuming that the Higgs particles are heavy enough to avoid $W'$ decays into them. An interesting feature of it is that, up to an overall normalization,  
the $W'$ boson has identical couplings to quarks and leptons as the SM $W$ boson. We refer to it as the ``meta-sequential" $W'$.

The most general scalar potential has many terms, but it is significantly simplified 
by imposing a $\mathbb{Z}_2$ symmetry (the bidoublet transforms into its charge conjugate). 
The lightest Higgs particle that is odd under this parity is stable, and could be a viable 
dark matter candidate. 
Whether or not the $\mathbb{Z}_2$ is exact, it leads to cascade decays of the $W'$ 
that give signatures with one or two electroweak bosons and two of these lightest odd particles (LOPs).

In Section \ref{model} we study the masses and couplings of the Higgs particles, and of the heavy gauge bosons.
Then, in Section \ref{decays}, we compute the branching fractions of the $W'$ and  $Z'$ bosons, and comment on various 
signatures arising from their cascade decays.
In Section \ref{signatures} we discuss the LHC phenomenology assuming that the LOPs escape the detector.
We summarize our results in Section \ref{conclusions}.

\section{\label{model} An $SU(2)\times SU(2)\times U(1)_Y$ model with odd Higgs sector}

Let us focus on a simple Higgs sector that breaks the $SU(2)_1 \times SU(2)_2 \times U(1)_Y$ gauge group down to $U(1)_{\rm em}$:  
a bidoublet complex scalar, $\Delta$, which has 0 hypercharge, and an $SU(2)_1$ doublet, $\Phi$. 
We take the SM quarks and leptons to be $SU(2)_2$ singlets.
The scalar and fermion gauge charges are shown in Table \ref{charges}.  \\[-1.cm]

\begin{table}[b] \renewcommand{\arraystretch}{1.2}
\begin{center}
\begin{tabular}{|c||c|c|c|c|} \hline
 & $\;  SU(3)_c \;  $ &  $\;   SU(2)_1 \;  $ & $\; SU(2)_2 \; $ &  $U(1)_Y$  \\ 
 \hline \hline 
 $\; \;  \;  \Delta \; \;  \; $ & 1 & 2 & $\bar 2$ & 0  \\ \hline
 $\Phi$     & 1 & 2 & 1 & $ \displaystyle \;\; \;   +\frac{1}{2} \; \; \;  $ \\ [0.3em]  \hline 
 \hline 
 $\;  Q_L  \, , \, L_L \; $        & $3 \, , \, 1$ & 2 & 1 & $ \displaystyle \; + \frac{1}{6}  \, , \, -\frac{1}{2}\; $\\ [0.3em] \hline 
 $\;  u_R \, , \, d_R \; $    & 3 & 1 & 1 & $ \displaystyle \;  \; +\frac{2}{3} \, , \,  -\frac{1}{3} \; \; $ \\ [0.3em]  \hline
 $e_R$    & 1 & 1 & 1 & $+1$ \\ \hline
\end{tabular}
\end{center}
\caption{Gauge assignments for the scalars ($\Delta$ and $\Phi$) and SM fermions.}
\label{charges}
\end{table} 

\subsection{Scalar spectrum}

We require the Lagrangian to be symmetric under the interchange $\Delta \leftrightarrow \tilde{\Delta}$,
where $\tilde{\Delta} $ is the charge conjugate of $\Delta$.
The most general renormalizable scalar potential exhibiting this $\mathbb{Z}_2$ symmetry and CP invariance  
is \cite{Barger:1980ix}  \\[-0.4cm]
\begin{eqnarray} \label{L}
V  &= &  m_\Phi^2 \, \Phi^{\dagger} \Phi +  \frac{\lambda_\Phi}{2}  \left( \Phi^{\dagger} \Phi \right)^2 
+ \left( m_\Delta^2  +  \lambda_0 \, \Phi^{\dagger}  \Phi  \right)  \, \Tr{ \left(\Delta^{\dagger} \Delta \right)} 
+ \frac{\lambda_\Delta}{2} \left[ \Tr{ \left(\Delta^{\dagger} \Delta \right) } \right]^2    
   \nonumber \\  [3mm]
&& \; -  \frac{\tilde{\lambda}}{2}  \left| \Tr{ \big(\Delta^{\dagger} \tilde{\Delta}  \big) } \right|^2
\, - \,  \left[  \frac{\tilde\lambda^\prime}{4}  \big( \!\Tr{ (\Delta^{\dagger} \Tilde{\Delta} ) } \big)^2 + {\rm H.c.} \right]   ~~.
\label{potential}
\end{eqnarray}

To avoid runaway directions, we impose  $\lambda_\Phi, \lambda_\Delta > 0$.
The  $\tilde{\lambda}$ and $\lambda_0$ quartic couplings must be real so that the potential is Hermitian.
The  $\tilde{\lambda}^\prime$ quartic coupling may be complex, but its phase can be rotated away by a 
redefinition of $\Delta$; we then take  $\tilde{\lambda}^\prime$ to be real without loss of generality. 

Canonical normalization of the $\tilde{\lambda}$  and $\tilde{\lambda}^\prime$ terms
would require an extra factor of $1/2$; we do not include it in order to simplify some equations below.
Other terms in $V$, such as  $\Tr{[ ( \Delta^{\dagger} \Delta )^2] }$, $\Tr{ ( \Delta^{\dagger} \Delta \tilde\Delta^{\dagger} \tilde\Delta) }$, or
$\Tr{ ( \Delta^{\dagger} \tilde\Delta \tilde\Delta^{\dagger} \Delta) }$,
would be redundant as they are linear combinations 
of the $\lambda_\Delta $, $\tilde{\lambda}$ and  $\tilde{\lambda}^\prime$ terms.
We recover the potential of Ref.~\cite{Barger:1980ix} using the identity
$ \Phi^{\dagger} (\Delta^{\dagger} \Delta + \tilde\Delta^{\dagger} \tilde\Delta)  \Phi = \Phi^{\dagger}  \Phi  \Tr{ \left(\Delta^{\dagger} \Delta \right)}   $. 

We also impose $m_\Delta^2 < 0$ so that $\Delta$ acquires a VEV. In addition, we need $m_\Phi^2 < 0$ or  $\lambda_0 < 0$
 such that $\Phi$ also acquires a VEV.
We are interested in the vacuum that preserves the $U(1)_{\rm em}$ and $\mathbb{Z}_2$ symmetries:
\beq
\langle \Delta \rangle = \frac{v_\Delta}{2}  \, {\rm diag} \left(1, 1 \right)  \;\;\;  ,  \;\;\; \;\;\;  \langle \Phi \rangle =  \frac{v_{\phi}}{\sqrt{2}} \,
\bigg( \begin{array}{c} 0 \\[-2mm] 1 \end{array} \bigg) ~~.
\label{vacuum}
\eeq
This vacuum is indeed a minimum of the potential for a range of parameters (discussed below).
The VEVs $v_\phi  > 0$ and $v_\Delta > 0$ are related to $m_\Phi^2$, $m_\Delta^2$, 
and the five quartic couplings by the extremization conditions: \\ [-0.9cm]
\begin{align}
\lambda_\star  v_\Delta^2 +  \lambda_0  v_\phi^2 & = - 2 m_\Delta^2
 ~~,  \nonumber \\[2mm]
\lambda_0 v_\Delta^2 +  \lambda_\Phi  v_\phi^2 & = -2 m_\Phi^2 ~~,
\label{extremization}
\end{align}
where we defined 
\beq
\lambda_\star \equiv \lambda_\Delta - \tilde\lambda -  \tilde\lambda^\prime  ~~.
\eeq

In terms of fields of definite electric charge, the scalars can be written as 
\begin{align}
& \Phi =  \left( \begin{array}{c}    \phi^+  \\ \displaystyle
\frac{1}{\sqrt{2}} \left(  v_{\phi} + \phi^{0}_r + i \phi^{0}_i \right)  \end{array} \right)  ~~,  \nonumber  \\[4mm]
&  \Delta =  \left( \begin{array}{cc} \eta^0 & \chi^+ \\ \eta^- & \chi^0  \end{array} \right)
 = \langle \Delta \rangle + \left( \begin{array}{cc}   \displaystyle
\frac{1}{\sqrt{2}}  \left( \eta^{0}_r+i\eta^{0}_i \right) \; & \chi^+ \\  
\eta^- &  \; \displaystyle
 \frac{1}{\sqrt{2}} \left( \chi^{0}_r+i\chi^{0}_i \right) \end{array} \right) ~~.
\label{components}
\end{align}
The charge conjugate state of the bidoublet is then
\beq
\tilde{\Delta} =  \sigma_2 \, \Delta^* \, \sigma_2  =  \left( \begin{array}{cc}
\chi^{0*} & -\eta^+ \\
-\chi^- & \eta^{0*}  \end{array} \right)   ~~.
\eeq

All odd fields under $\mathbb{Z}_2$ (which cannot mix with even fields, and thus are already in the mass eigenstate basis) 
are collected in 
\beq
\Delta - \tilde{\Delta} =   \left( \begin{array}{cc}  H^0 + i A^0  & \sqrt{2} \, H^+ \\[2mm] \sqrt{2} \, H^-  & - H^0 + i A^0   \end{array} \right) ~~,
\eeq
where the physical states consist of  a CP-even scalar ($H^0$),  a CP-odd scalar ($A^0$),  and a charged scalar ($H^\pm$).
These are related to the $\eta$ and $\chi$ fields by
\begin{align}
& A^0 = \frac{1}{\sqrt{2}} \left( \eta^0_i  + \chi^0_i   \right)   ~~,  \nonumber \\[2mm]
& H^0 = \frac{1}{\sqrt{2}} \left( \eta^0_r - \chi^0_r  \right)   ~~,    \nonumber \\[2mm]
& H^\pm = \frac{1}{\sqrt{2}} \left( \eta^\pm + \chi^\pm  \right)   ~~.
\end{align}
At tree-level, the  $\mathbb{Z}_2$-odd scalars have masses  given by \\[-1cm]
\begin{align}
 & M_A =  \sqrt{  2 \tilde\lambda^\prime }\; v_{\Delta}   ~~, \nonumber \\[2mm]
 & M_{H^+} = M_{H^0}  =  \sqrt{ \tilde{\lambda} + \tilde\lambda^\prime } \;\, v_{\Delta} ~~.
 \label{oddmasses}
\end{align} 

The are two remaining scalars not eaten by the gauge bosons. These are $\mathbb{Z}_2$-even, CP-even, and neutral; their
mass-squared matrix
in the  $(\chi^0_r + \eta^0_r)/\sqrt{2}$ , $\phi^0_r$  basis is \\ [-0.3cm]
\beq 
\mathcal{M}^2_{\rm even} = \left(
\begin{array}{lr}     
\;\;  \lambda_\star \, v_{\Delta}^2   &   \;\;\; \lambda_0  \, v_\phi \, v_\Delta   \\ [4mm] 
 \lambda_0  \, v_\phi \, v_\Delta   \;\;\;  &  \lambda_\Phi \,  v_\phi^2  \;\;
\end{array}
\right)  ~~.
\eeq
The $\mathbb{Z}_2$-even physical scalars, 
 \begin{align}
h^0 &=  \phi^0_r  \, \cos{\alpha_h} - \frac{1}{\sqrt{2}} \left( \chi^0_r + \eta^0_r  \right)  \, \sin{\alpha_h}   ~~, \nonumber \\[2mm]
H^{\prime \, 0} &= \phi^0_r \, \sin{\alpha_h}  +  \frac{1}{\sqrt{2}} \left( \chi^0_r + \eta^0_r  \right)  \, \cos{\alpha_h}        ~~,
 \end{align}
have the following squared masses:
  \beq \label{mixedMasses}
 M_{h, H^\prime}^2  = \frac{1}{2} 
 \left(   \lambda_\star  v_{\Delta}^2  + \lambda_\Phi v_\phi^2 
 \mp \sqrt{ \left(  \lambda_\star  v_{\Delta}^2  - \lambda_\Phi v_\phi^2  \right)^2 + 4 \lambda_0^2 \,  v_\phi^2 v_{\Delta}^2 } \; \right) ~~.
 \eeq
 The mixing angle $\alpha_h$ satisfies 
 \beq \label{alpha}
\tan 2\alpha_h =  \frac{2\lambda_0 v_{\Delta} v_\phi}{\lambda_\star v_{\Delta}^2 -  \lambda_\Phi  v_\phi^2 }  ~~.
 \eeq
  
 The necessary and sufficient conditions for the vacuum (\ref{vacuum}) to be a minimum of the potential are \\ [-1.1cm]
 \begin{align}
 \tilde\lambda^\prime & > {\rm Max } \{ - \tilde\lambda , 0 \}   ~~,
\nonumber \\[1mm]
 \lambda_\star \lambda_\Phi  &  >  \lambda_0^2  ~~, 
\nonumber \\[2mm]
  \lambda_\Phi  |m_\Delta^2|  & >  - \lambda_0  m_\Phi^2 ~~, 
\nonumber \\[2mm]
 \lambda_0   |m_\Delta^2|   & >  -  \lambda_\star m_\Phi^2   ~~;
\end{align}
these follow from imposing that all physical scalars have positive squared masses [see 
Eqs.~(\ref{oddmasses}) and (\ref{mixedMasses})], and 
that the extremization conditions (\ref{extremization})  have solutions.
 
All above results are valid for any $v_\phi/v_\Delta$. 
The agreement between SM predictions and the data suggests that the Higgs sector is near the decoupling limit
$v_\phi^2 \ll v_\Delta^2$; adopting this limit,
we can analyze the spontaneous symmetry breaking in two stages. The first one is $SU(2)_1\times SU(2)_2\times U(1)_Y \to SU(2)_W \times U(1)_Y $ at the scale $v_\Delta$. The effective theory below $v_\Delta$ consists of  the SM (with the Higgs doublet $\Phi$) plus 
an $SU(2)_W$-triplet of heavy gauge bosons ($W^{\prime \pm} , Z'$), and five of the scalar degrees of freedom from $\Delta$:
four $\mathbb{Z}_2$-odd scalars combined into an  $SU(2)_W$-triplet  ($H^\pm, H^0$) and a singlet ($A^0$), and 
a $\mathbb{Z}_2$-even singlet ($H^{\prime \, 0}$).

The second stage of symmetry breaking is the SM one: $SU(2)_W \times U(1)_Y \to U(1)_{\rm em}$ at the weak scale $v_\phi \approx 246$ GeV.
The lightest CP-even scalar, $h^0$, represents the recently discovered Higgs boson, because its couplings are the same as the SM ones 
up to small corrections of order $v_\phi^2 / v_\Delta^2$.
Its mass is given by 
\beq
M_h  =  v_\phi  \left( \lambda_\Phi - \frac{\lambda_0^2}{ \lambda_\star}  \right)^{\! 1/2} 
\left[ 1 -\frac{\lambda_0^2 \, v_\phi^2}{ 2 \lambda_\star^2 \, v_\Delta^2}  + O\!\left( v_\phi^4/v_\Delta^4 \right) \right] ~~,
\eeq
and should be identified with the measured Higgs mass, near 126 GeV.
The $H^{\prime \, 0}$ even scalar has the same couplings as the SM Higgs except for an overall suppression by 
\beq \label{alpha-approx}
\sin \alpha_h =  \frac{\lambda_0 \, v_\phi}{  \lambda_\star v_\Delta} + O\!\left( v_\phi^3/v_\Delta^3 \right) ~~,
\eeq
and is significantly heavier:
\beq
M_{H^\prime} = \sqrt{\lambda_\star} v_\Delta + O\!\left( v_\phi^2/v_\Delta \right)  ~~.
\eeq
Consequently, its dominant decay modes are $W^+W^-$ and $ZZ$.

The odd scalars, $H^\pm$, $H^0$,  $A^0$, couple exclusively to gauge bosons and scalars, and only in pairs. 
The lightest of them is stable, and a component of dark matter.
$A^0$ is naturally the lightest odd particle (LOP). 
because in the $\tilde\lambda^\prime \to 0 $ limit the symmetry is enhanced: $A^0$ becomes the Nambu-Goldstone 
boson of a global $U(1)$ symmetry acting on $\Delta$.
We note, however, that $H^0$  could also be the LOP (for $\tilde\lambda^\prime  > \tilde\lambda$)
and a viable dark matter candidate. Even though it is part of an $SU(2)_W$  triplet that is degenerate at tree-level, 
electroweak loops split the $H^\pm$ and $H^0$ masses \cite{Dodelson,Cirelli:2005uq}.

In what follows we will assume that $A^0$ is the LOP.
The heavier odd scalars then decay as follows: $H^\pm \to W^\pm A^0$, $H^0 \to Z A^0$. Even when 
these two-body decays are kinematically forbidden, the three-body decays through an off-shell $W^\pm$ or $Z$
are the dominant ones. 
Other channels are highly suppressed, either kinematically ($H^+\to \pi^+ \pi^0 H^0$ and $H^+\to \ell^+ \nu H^0$) 
or by loops ($H^0 \to \gamma A^0$ and the CP-violating $H^0 \to h^0 A^0$).

\subsection{Meta-sequential $W'$ boson}

The kinetic terms for the $\Phi$ and $\Delta$ scalars,
\beq
(D_{\mu}\Phi)^\dag  D_{\mu}\Phi+\Tr \left[ (D_{\mu}\Delta)^\dag  D_{\mu}\Delta  \right]  ~~,
\eeq
involve the covariant derivative 
\beq
D_{\mu} = \partial _{\mu} -i g_Y Y B_\mu- i g_1 \vec{T}_1 \cdot \vec W_{1 \mu} - i g_2 \vec T_2 \cdot \vec W_{2 \mu}  ~~,
\eeq
with $T_{1,2}  = \sigma_{1,2}/2 $;  notice that $\vec{T}_2$ acts from the right on the bidoublet:   
$\vec{T}_2 \cdot \Delta = -  \Delta \cdot \vec \sigma /2$.
After symmetry breaking,  the electrically-charged gauge bosons acquire mass terms:
\beq
\frac{v_\phi^2}{4}  \,  g_1^2 \, W^+_{1 \mu} W^{-\mu}_1 + 
\frac{v_\Delta^2}{4}  \left(  g_1  W^+_{1 \mu} -  g_2  W^+_{2 \mu} \right)  \left(  g_1  W^{-\mu}_1 -  g_2  W^{-\mu}_2 \right)   ~~.
\eeq
Diagonalizing them gives the physical charged spin-1 states,
\begin{align}
W_\mu & = W_{1 \mu} \cos{\theta} + W_{2 \mu} \sin{\theta}    ~~,
\nonumber \\
W^\prime_\mu & =  - W_{1 \mu} \sin{\theta}  + W_{2 \mu}  \cos{\theta}~~,
\label{W-eigenstate}
\end{align}
with the following mixing angle, $0 \leq \theta \leq \pi/2$:
 \beq \label{theta}
\tan 2\theta =  \frac{2 g_1 g_2 \, v_{\Delta}^2 }{ \left( g_2^2 - g_1^2 \right) v_{\Delta}^2 -  g_1^2 v_\phi^2 }  ~~.
 \eeq
The masses of the $W$ and $W^\prime$ bosons are
\beq
M_{W,W'} = \frac{1}{2\sqrt{2}} \left[ \left( g_2^2 + g_1^2 \mp  \frac{2 g_1 g_2 }{\sin 2\theta} \right) v_{\Delta}^2 + g_1^2 v_\phi^2 \right]^{1/2}   ~~.
\label{exact-Wmass}
\eeq 

Given that the left-handed quarks and leptons transform as doublets only under $SU(2)_1$,
their couplings to the $W$ and $W^\prime$ bosons are proportional to the respective coefficients of $W_{1\mu}$ in
Eqs.~(\ref{W-eigenstate}).  
The measured $W$ coupling to fermions gives a value for the $SU(2)_W$ gauge coupling of 
$g = \sqrt{4 \pi \alpha}/s_W \approx 0.652$, where the electromagnetic coupling constant and the weak mixing angle are evaluated at the $M_Z$ scale: $\alpha \equiv \alpha (M_Z) \approx 1/127.9$ and
$s_W \equiv \sin{\theta_W} \approx \sqrt { 0.231}$.
In terms of the parameters of this model, the $SU(2)_W$ gauge coupling can be expressed as
\beq
g_1 \cos\theta = g ~~.
\label{weak-coupling}
\eeq
The $W'$ coupling to quarks and leptons, derived from Eq.~(\ref{W-eigenstate}) and Table I, is then
\beq
- g_1 \sin\theta = - g \tan\theta  ~~.
\label{Wprime-coupling}
\eeq
Thus, $\tan\theta$ determines completely the  tree-level couplings of $W'$ to SM fermions.
Imposing a perturbativity condition on the $SU(2)_1\times SU(2)_2$ gauge couplings,
$g_{1,2}^2/(4\pi)  \lesssim 1$, and using Eq.~(\ref{weak-coupling}) we find that 
\beq
0.2 \lesssim  \tan{\theta}  \lesssim  5  ~~.
\eeq

In the particular case of $\tan\theta =1$, the couplings of $W'$ to fermions are identical (at tree level) 
to those of the $W$. This is usually referred to as the sequential $W'$ boson, and is a common benchmark model for $W'$
searches at colliders. The most recent limit on the mass of a sequential $W'$ at CMS, using 20 fb$^{-1}$ of 8 TeV data, is 3.8 TeV \cite{CMS:2013rca}, assuming that $W'$ can decay only into SM fermions. Note that the relative sign 
in Eqs.~(\ref{weak-coupling}) and (\ref{Wprime-coupling}) implies constructive interference between the $W$ and $W'$ amplitudes that contribute to processes constrained by $W'$ searches at the LHC.
In the next sections we will focus on the region $0.2<\tan{\theta}<1$, where the LHC limits are relaxed.
Given that the $W'$ boson in this model has couplings to fermions 
proportional to the SM $W$ ones (by an overall factor of $-\tan\theta$), we refer to it as a ``meta-sequential $W'$ boson". 

The above results are valid for any $v_\phi/v_\Delta$. It is instructive to expand these results in 
powers of $(v_\phi/v_\Delta)^2 \ll 1$.
The $W'$ coupling to fermions, relative to the $W$ one is
\beq
\tan\theta =  \tan{\theta_0} \left( 1 - \frac{ v_\phi^2}{ v_\Delta^2}  \, \cos^2\!\theta_0 \right)  + O\!\left( v_\phi^4/v_\Delta^4 \right)  ~~,
\label{tan-theta}
\eeq
where we defined 
\beq
\tan{\theta_0} \equiv \frac{g_1}{g_2}   ~~.
\eeq
For $v_\phi^2 \ll v_\Delta^2$, the values of $\tan{\theta_0}$ span essentially the same range as $\tan{\theta} $.
The $W$ and $W'$ masses, given in  Eq.~(\ref{exact-Wmass}), have simple expressions to leading 
order in  $v_\phi/v_\Delta$: \\[-5mm]
\begin{align}
& M_{W} = \frac{g_2}{2} \, v_\phi \sin\theta_0  \left[ 1 -   \frac{ v_\phi^2}{ 2v_\Delta^2}  \, \sin^4\!\theta_0 + O\!\left( v_\phi^4/v_\Delta^4 \right) \right]  ~~,
\label{W-mass}
 \\[2mm]
& M_{W'} = \frac{g_2 \, v_\Delta}{2 \cos\theta_0} \left[ 1 +   \frac{ v_\phi^2}{ 2v_\Delta^2}  \, \sin^4\!\theta_0 + 
\frac{ v_\phi^4 }{ 8 v_\Delta^4 } \left( 4 \cot^2\!\theta_0 -1\right) \, \sin^8\!\theta_0 + 
O\!\left( v_\phi^6/v_\Delta^6 \right) \right]  ~~.
\label{Wprime-mass}
\end{align}

\vspace*{3mm}

The low-energy charged current interactions are mediated in this model by both $W$ and $W'$ exchange.
Consequently, the Fermi constant is related to our parameters by\\[-5mm]
\begin{align}
  4\sqrt{2} G_F &  =  \frac{(g_1\cos\theta)^2}{M_W^2} + \frac{(g_1\sin\theta)^2}{M_{W'}^2}  
\nonumber \\[2mm]
& = \frac{g^2}{M_W^2} \, \left[ 1 + \frac{ v_\phi^2}{ v_\Delta^2} \, \sin^4\!\theta_0  + 
O\!\left( v_\phi^4/v_\Delta^4 \right)   \right]    ~~,
\label{GFermi}
\end{align}

\vspace*{2mm}

\noindent
where we used Eq.~(\ref{weak-coupling}), which defines $g$ as the tree-level $W$ coupling to leptons and quarks.
This shows that the measurements of the weak coupling in  low-energy processes
and  in collider processes involving $W$ bosons
should agree up to tiny corrections of order $(v_\phi/v_\Delta)^2 \sin^4\!\theta_0$.
Defining the weak scale $v \approx 246$ GeV through $G_F = 2^{1/2} v^{-2}$, and using Eq.~(\ref{W-mass}), 
we obtain the relation between the $\Phi$ VEV and the weak scale
\beq
v = v_\phi   \left[ 1 -   \frac{ v_\phi^2}{ v_\Delta^2}  \, \sin^2\!\theta_0 + O\!\left( v_\phi^4/v_\Delta^4 \right) \right]  ~~.
\eeq

\vspace*{2mm}

\subsection{$Z'$ mass and couplings}

Electrically-neutral gauge bosons also acquire  mass terms in the vacuum (\ref{vacuum}):
\beq
\frac{v_\phi^2}{8}   \left(  g_1 \, W^3_{1 \mu} - g_Y B_\mu \right)^2 + 
\frac{v_\Delta^2}{8}  \left(  g_2  W^3_{2 \mu} - g_1  W^3_{1 \mu}  \right)^2  ~~.
\label{Zmass-terms}
\eeq
It is convenient to diagonalize these in two steps. First, we 
define some intermediate fields denoted with hats:
\begin{align}
\hat{Z}'_{\mu} &= W_{2 \mu}^3  \cos{\theta_0}   - W_{1 \mu}^3   \sin{\theta_0}  ~~,
\nonumber \\[2mm]
\hat{Z}_{\mu} &=  \left( W_{2 \mu}^3 \sin{\theta_0} + W_{1 \mu}^3  \cos{\theta_0} \right) \cos{\hat\theta_W}  - B_\mu  \sin{\hat\theta_W}   ~~,
\end{align}
where the angle $\hat\theta_W$ is defined in terms of coupling ratios:
\beq
 \tan{\hat\theta_W} = \frac{g_Y}{g_2  \sin{\theta_0} } ~~.
\eeq
The gauge boson orthogonal to $\hat{Z}_{\mu}$ and  $\hat{Z}'_{\mu}$ is the photon 
($A_\mu = W_{1 \mu}^3 \cos{\theta_0}  \sin{\hat\theta_W} + B_\mu  \cos{\hat\theta_W} $), already in the physical eigenstate.
The measured electromagnetic coupling, $e =  \sqrt{4 \pi \alpha} \approx 0.313$, is related to the original gauge couplings through
\beq
\label{edef}
g_Y \cos\hat\theta_W  = e ~~.
\eeq
The mass-squared matrix for $\hat{Z}_\mu$ and $\hat{Z}_\mu^\prime$ takes the form \\[-3mm]
\beq
\mathcal{M}^2_{Z} = \frac{g_2^2}{4}\sin^2\!\theta_0 \left(\begin{array}{cc}  \displaystyle
 \frac{  v_\phi^2  }{\cos^2\!\hat\theta_W}  \;\;  & \;\;  \displaystyle -  \frac{v_\phi^2  \tan\theta_0}{\cos\hat\theta_W}  \\[6mm]
  \displaystyle -  \frac{v_\phi^2  \tan\theta_0}{\cos\hat\theta_W}  \; \; 
  &  \displaystyle \;\;\;\;  \frac{4v_\Delta^2}{\sin^2\!2\theta_0}  + v_\phi^2 \tan^2\!\theta_0   \end{array}\right)    ~~.
\eeq

\vspace*{2mm}

In the second step, we rotate $\hat{Z}_{\mu}$ and  $\hat{Z}'_{\mu}$ by an angle $\epsilon_Z$, given by \\[-3mm]
 \beq \label{epsilon}
\tan 2\epsilon_Z =  \frac{v_{\phi}^2 \sin 2\theta_0 \,  \sin^2\!\theta_0 \, \cos\hat\theta_W }{  v_{\Delta}^2 \cos^2\!\hat\theta_W  + v_\phi^2 \sin^4\!\theta_0 \left( \cos^2\!\hat\theta_W  - \cot^2\!\theta_0 \right) }  ~~,
 \eeq
 
\vspace*{2mm}
 
 \noindent
in order to obtain the mass eigenstate $Z$ and $Z^\prime$ bosons:\\[-9mm]
\begin{align}
Z_{ \mu} &= \hat{Z}_{ \mu} \cos{\epsilon_Z}  + \hat{Z}'_{\mu} \sin{\epsilon_Z}   ~~,  \nonumber \\[1mm]
Z'_{\mu} &=- \hat{Z}_{ \mu} \sin{\epsilon_Z}  +  \hat{Z}'_{\mu} \cos{\epsilon_Z}  ~~.
\end{align}
The masses of the heavy neutral spin-1 particles are \\[-3mm]
\beq
M_{Z,Z'} = \frac{g_2}{2\sqrt{2}} \left[ \frac{v_\Delta^2 }{\cos^2\!\theta_0} + v_\phi^2 \sin^2\!\theta_0  
\left(  \frac{1}{\cos^2\!\hat\theta_W}  +  \tan^2\!\theta_0 \mp \frac{ 2\tan\theta_0}{\sin 2\epsilon_Z \,  \cos\hat\theta_W} \right) \right]^{1/2}  ~~.
\label{exact-Zmass}
\eeq

\vspace*{4mm}

The tree-level results (\ref{Zmass-terms})-(\ref{exact-Zmass}) have been obtained without approximations.
Expanding now in $v_\phi^2/v_\Delta^2$, we find \\[-5mm]
\begin{align}
& M_{Z} = \frac{g_2 v_\phi \sin\theta_0 }{2\cos\hat\theta_W} \,  \left[ 1 -   \frac{ v_\phi^2}{ 2v_\Delta^2}  \, \sin^4\!\theta_0 + O\!\left( v_\phi^4/v_\Delta^4 \right) \right]  ~~,
\label{Z-mass}
 \\[2mm]
& M_{Z'} = \frac{g_2 \, v_\Delta}{2 \cos\theta_0} \left[ 1 +   \frac{ v_\phi^2}{ 2v_\Delta^2}  \, \sin^4\!\theta_0 + 
\frac{ v_\phi^4 }{ 8 v_\Delta^4 } \left( 4 \frac{\cot^2\!\theta_0}{\cos^2\!\hat\theta_W} -1\right) \, \sin^8\!\theta_0 + 
O\!\left( v_\phi^6/v_\Delta^6 \right) \right]  ~~.
\label{Zprime-mass}
\end{align}

\vspace*{2mm}

\noindent
The original five parameters from the gauge sector ($g_1,g_2,g_Y,v_\phi, v_\Delta$) can be traded for 
three observables ({\it e.g.}, $e, s_W, M_W$) and two parameters that can be measured once the $W'$ or $Z^\prime$
boson is discovered  ($M_{W'}, \tan\theta$), using Eqs.~(\ref{tan-theta}), (\ref{W-mass}), (\ref{Wprime-mass}), (\ref{edef})
and 
\begin{align}
s_W = \sin \hat\theta_W \,  \left[ 1 -   \frac{ v_\phi^2}{ v_\Delta^2}  \, \sin^2\!\theta_0 \cos^2\!\theta_0
+ O\!\left( v_\phi^4/v_\Delta^4 \right) \right]  ~~.
\end{align}

Eqs.~(\ref{Z-mass}) and (\ref{Zprime-mass}), combined with the above equation, 
show that the tree-level relation $M_Z c_W = M_W$, where
$c_W \equiv \cos{\theta_W}$, is satisfied only up to corrections of order $v_\phi^2/v_\Delta^2$.
Furthermore, the $Z$ couplings to fermions are modified at order $v_\phi^2/v_\Delta^2$ compared to the 
SM. Thus, the current agreement between electroweak measurements and the SM imposes
an upper limit on  $v_\phi^2/v_\Delta^2$, or equivalently, a lower limit on the $W'$ mass for a fixed 
$\tan\theta$.
The lower limit at the 95\% CL given by the global fit performed in Ref.~\cite{Cao:2012ng} increases from 
$M_{W'} \gtrsim 600$ GeV for $\tan\theta = 0.2$, to  $M_{W'} \gtrsim 2$ TeV for $\tan\theta =1 $ ({\it i.e.}, sequential $W'$).

The relative mass splitting between $W'$ and $Z'$ is very small:
\beq
\frac{M_{Z'}}{M_{W'}} -1 = \frac{ s_W^2}{2 c_W^2}  \tan^2\!{\theta} \, \left(\frac{M_W}{M_{W'}} \right)^{\!4}  + O\!\left( M_W^6/M_{W'}^6 \right)  ~~,            
\eeq
which is less than $6 \times 10^{-6}$ for $M_{W'} > 1$ TeV and $\tan{\theta} < 1$.
This implies that the $W'$ mass and $\tan\theta$ will be constrained
by both $Z'$ and $W'$ searches. 
The $Z^\prime$ interacts with the left-handed fermion doublets, with a coupling given by 
$g \tan\theta \; T^3$ plus corrections of order $v_\phi^2/v_\Delta^2$ that are different for quarks and leptons..
The $Z^\prime$ couplings to $SU(2)_W$ singlets are suppressed by  $v_\phi^2/v_\Delta^2$.

\section{\label{decays} $W'$ and $Z'$ decays}\setcounter{equation}{0}

The new gauge bosons interact with SM fermions and gauge bosons, as well as with the Higgs particles. 
Usually, resonance searches for new gauge bosons rely on
sizable branching fractions of the $W'$ and $Z'$ decays into SM fermions.
However, if the scalars are lighter than the vector bosons than the decays into SM fermions may be suppressed.
In our model, the left-handed fermion doublets transform under $SU(2)_1$, while all fermions are singlets 
under $SU(2)_2$. Thus, the $W'$ and $Z'$ couplings to fermions 
are induced through mixing with the $W$ and $Z$, so that for small $\tan{\theta}$
decays to heavy scalars become important. 

Neglecting corrections of $O(v_\phi^2/v_\Delta^2)$, the $W'$ and $Z'$ coupling to fermion doublets is given by
$g \tan{\theta}$.  The partial widths for decays to leptons (without summing over flavors) 
\beq
\Gamma(W' \rightarrow \ell \nu) \approx  2\Gamma(Z' \rightarrow \ell^+ \ell^-) \approx  \frac{\alpha}{6 s^2_W}  \tan^2\!{\theta}  \; M_{W'}    ~~, 
\eeq
are suppressed for $0.2<\tan{\theta}<1$.
By contrast, the $W'$ and $Z'$ couplings to pairs of odd Higgs particles are enhanced  by $1/\tan{\theta}$:
\begin{align}
g_{W'H^{\pm}A^0} = g_{Z'H^{0}A^0} &= \frac{ g }{  \sin{ 2\theta}  } ~~, \nonumber \\[2mm]
g_{W'H^{\pm}H^0} = g_{Z'H^{+}H^-}&= \frac{g}{\tan{2\theta} }  ~~ ,
\end{align}
where we ignored corrections of order $v_\phi^2/v _\Delta^2$.
These couplings lead to the following partial widths: 
\begin{align}
\Gamma(W' \!\rightarrow H^\pm A^0)  &\approx \Gamma(Z'  \!\rightarrow H^0 A^0) \approx  \frac{\alpha M_{W'}}{12 s^2_W \sin^2{\!2\theta}}  
\left(1 - 2 \frac{M_{H^+}^2  \!+ \! M_{A^0}^2}{M_{W'}^2} +\frac{(M_{H^+}^2  \!- \! M_{A^0}^2)^2}{M_{W'}^4}\right)^{\! 3/2}  , \nonumber \\[2mm]
\Gamma(W'  \!\rightarrow H^\pm H^0) &\approx \Gamma(Z'  \!\rightarrow H^+ H^-) \approx  
\frac{\alpha M_{W'}}{12 s^2_W \tan^2{\!2\theta} }
\left(1 - 4 \frac{M_{H^+}^2}{M_{W'}^2} \right)^{\! 3/2}   ~~.
 \end{align}
The $W'$ can also decay into $WZ$ and $Wh^0$ final states, but these partial widths 
are suppressed by $v_\phi^4/v_\Delta^4$.

\begin{figure}[t]
\includegraphics[width=3.18in]{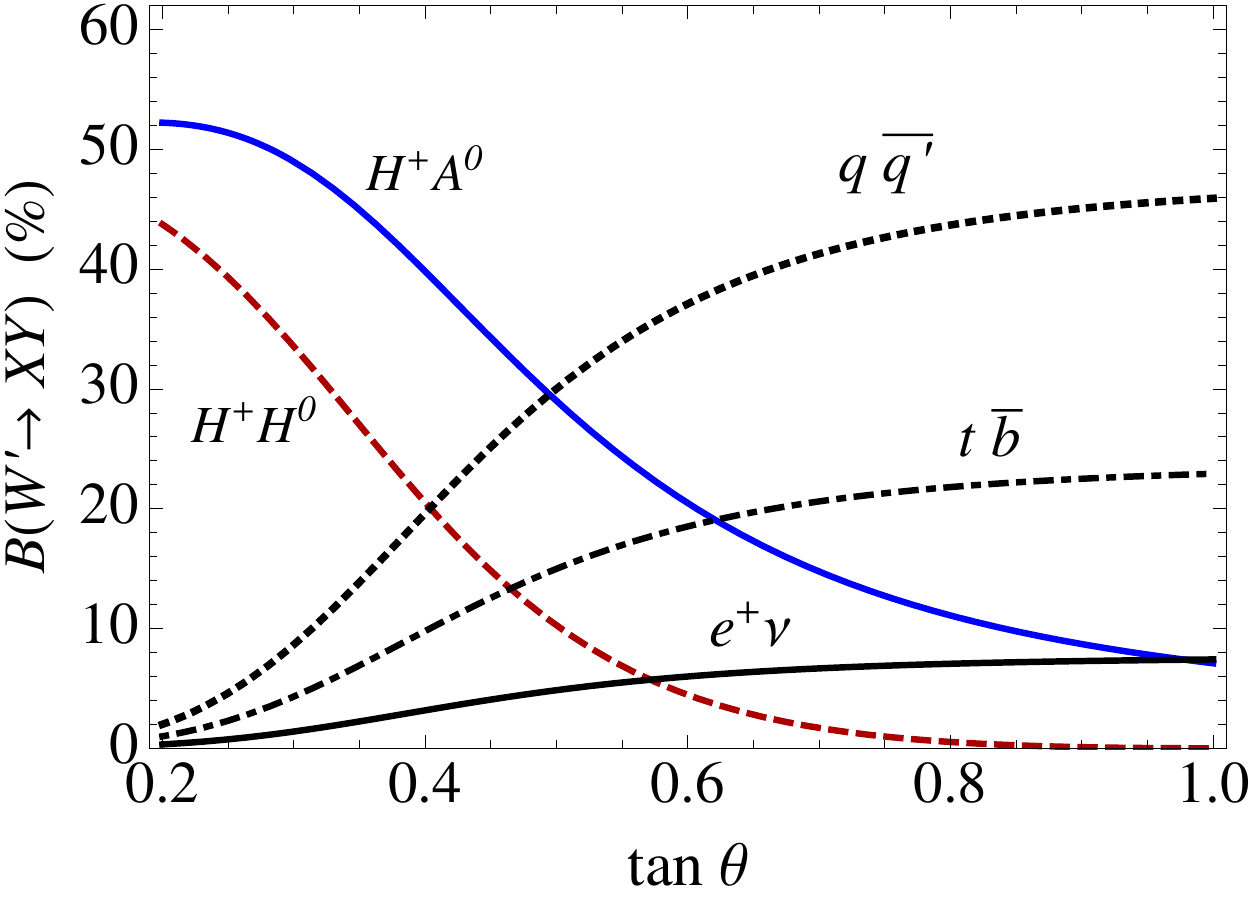} \hspace{0.cm}
\includegraphics[width=3.18in]{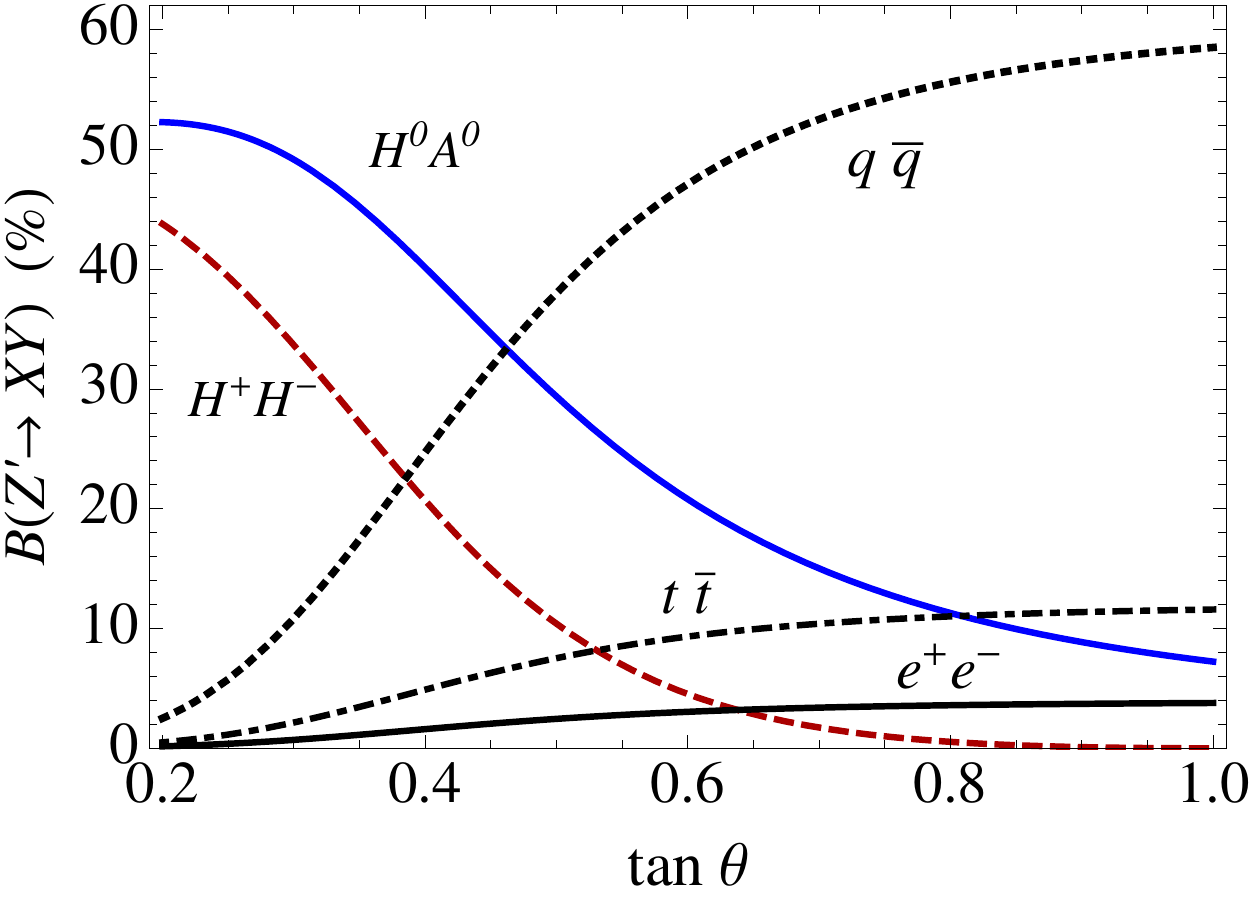}
\caption{\label{br}$W'$ and $Z^\prime$ branching fractions as a function of mixing angle, for $M_{W'} = 3$ TeV, $M_{H^+} = 300$ GeV, $M_{A} = 200$ GeV.} 
\end{figure}

Figure \ref{br} shows the branching fractions of the $W'$ and $Z'$ as a function of $\tan{\theta}$ for the dominant channels. As a benchmark point, we have used $M_{W'} = 3$ TeV, $M_{H^+} = 300$ GeV and  $M_{A} = 200$ GeV (as shown in 
Section II, $M_{W'} = M_{Z'}$ and  $M_{H^+} =  M_{H^0}$ to a good accuracy).
For $\tan{\theta} \lesssim 0.4$, the $W'$ decays dominantly to pairs of odd Higgs particles.
It is important to investigate collider signatures of these decays.

The heavier odd scalars decay into the LOP (taken to be $A^0$) and an electroweak boson, so that $W'$ and
$Z'$ can each undergo two cascade decays: $W' \!\! \rightarrow H^+ A^0 \to W A^0 A^0$,
$W' \!\! \rightarrow H^+ H^0 \to W^+ A^0  Z A^0$  (see Figure 2), 
and $Z' \!\! \rightarrow H^0 A^0 \to Z\,  A^0 A^0$,  $Z' \!\! \rightarrow H^+ H^- \to W^+ A^0 \, W^- A^0$.

If the $\mathbb{Z}_2$ symmetry discussed in Section \ref{model} is exact, then $A^0$ is a component of dark matter.
We will not explore here the constraints on the parameter space from the upper limit on relic density,
nor from direct detection experiments (nuclear scattering would occur through 
Higgs exchange and gauge boson loops);
these constraints can be in any case relaxed by allowing a tiny $\mathbb{Z}_2$ violation in the scalar potential. 
While an in-depth exploration of this model as an explanation for dark matter is left for future work, we note that it shares many features with inert doublet \cite{Barbieri:2006dq} and minimal dark matter scenarios  \cite{Cirelli:2005uq}.

\begin{figure}[t]
\begin{center}
\includegraphics[trim = 25mm 177mm 40mm 28mm, clip, height=3.1cm]{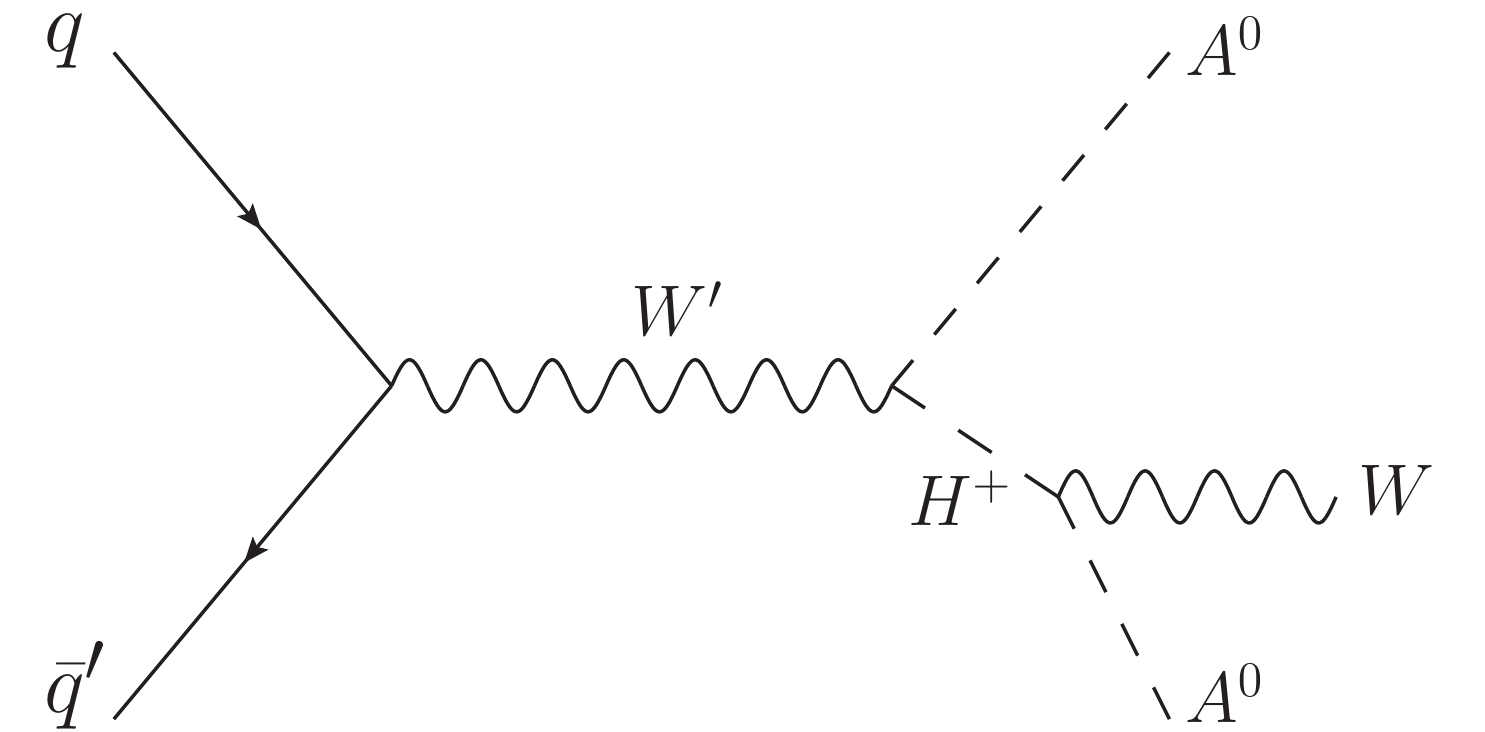}
\includegraphics[trim = 25mm 75mm 40mm 130mm, clip, height=3.1cm]{wp_diagrams.pdf}
\caption{\label{diagrams} $W'$ production and cascade decays through odd Higgs particles.}
\end{center}
\end{figure}

The possibility that the   $\mathbb{Z}_2$ symmetry is violated by terms in the scalar potential of the type \\[-10mm]
\beq
\Tr{ \big(\Delta^{\dagger} \tilde\Delta \big)} \;\; , \;\; \Phi^\dagger \tilde\Delta^{\dagger} \tilde\Delta \Phi  \;\; ,  \;\;
\Tr{ \big(\Delta^{\dagger} \Delta \Delta^{\dagger} \tilde\Delta \big)}     \;\; ,  \;\;
\Tr{ \big(\Delta^{\dagger} \Delta\big) } \Tr{\big( \Delta^{\dagger} \tilde\Delta \big)}    ~~, 
\label{ops-violation}
\eeq
is also worth considering.
The weak-triplet scalar ($H^\pm, H^0$) as well as the singlet $A^0$
would mix with the $\Phi$ doublet, allowing direct two-body decays of $A^0$, $H^0$ and $H^\pm$ to SM particles. 
Furthermore, the three CP-even neutral scalars ($H^0, h, H'$) would then mix, so that $W'$ and $Z'$ decays 
involving the SM-like Higgs boson are possible. These include $W' \to H^+ h^0$ with $H^+ \to t\bar b$ 
(this channel is analyzed in \cite{poster}), 
as well as $W' \to H^+ h^0 \to W^+ A^0 h^0$ and $Z' \to h^0 A^0$ 
with $A^0 \to b \bar b$ (or $t \bar t$ if kinematically allowed). There are, however, various constraints on 
deviations from the SM Higgs couplings, implying that the $\mathbb{Z}_2$  violating mixing is small, so that 
we expect that the above final states have relatively small branching fractions.
 
It is also interesting to consider the intermediate case, 
where the violation of $\mathbb{Z}_2$ is very small, {\it i.e.}, the coefficients of 
the operators (\ref{ops-violation}) are much less than one.
In that case all $W'$ and $Z'$ cascade decays through the odd Higgs particles proceed as 
before, but the $A^0$ would decay to a pair of heaviest fermions of mass below $M_A/2$.
This leads to a variety of noteworthy final states: $W' \!\! \rightarrow WZ+4b$, $Z' \!\! \rightarrow Z+4b$, or 
$W' \!\! \rightarrow WZ t\bar t t\bar t$, $Z' \!\! \rightarrow Zt\bar tt\bar t$, etc.
For a range of parameters, the decays of $A_0$ may be displaced but still within the detector, 
leading to potentially confusing events.
In what follows we will consider only the case where $A^0$ is stable enough to escape the detector.

\begin{figure}[b]
 \hspace{-0.2cm}\includegraphics[width=4.1in]{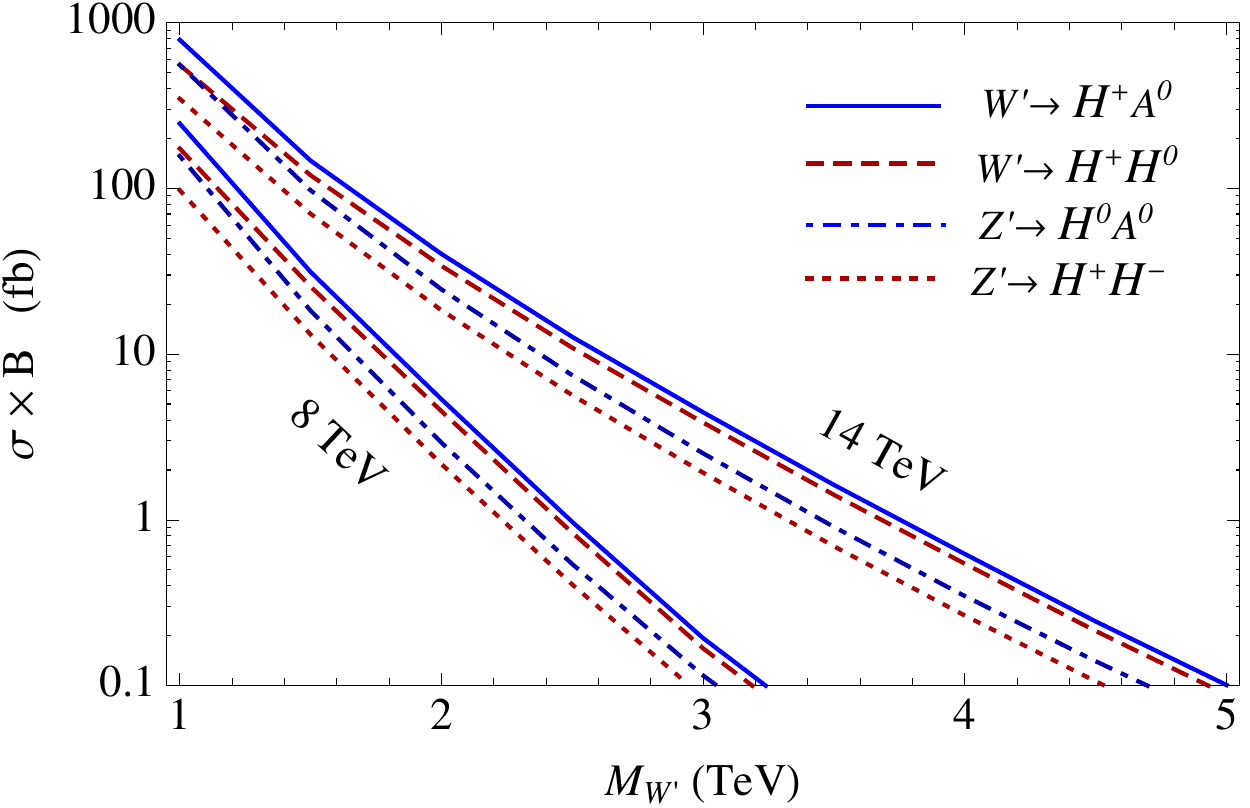}
\caption{\label{xsec} Leading-order cross sections times branching fractions 
for the processes $p p \rightarrow W' \!\! \rightarrow H^+ A^0, \, H^+ H^0$
and $p p \rightarrow Z' \rightarrow H^0 A^0, \,  H^+ H^-$ at $\sqrt{s} = 8 $ TeV and 14 TeV.
 We have chosen $\tan{\theta} = 1/4$, $M_{H^+} = 300$ GeV and $M_{A} = 200$ GeV.}
\end{figure}

\setcounter{equation}{0}
\section{\label{signatures}LHC signatures with stable $A^0$}

At the LHC, the $W'$ boson would be mainly produced in the $s$ channel from quark-antiquark initial state, even for small $\tan\theta$.  In the narrow width approximation, the leading-order cross section for  $W^\prime$ production followed by 
decay into  $H^+ A^0$ or $H^+ H^0$ is
\beq 
\sigma(p p \rightarrow W' \rightarrow H^+ A^0 , H^+ H^0) \approx  \frac{\alpha  \tan^2{\theta}  }{24 s^2_W  \, s} 
w(M_{W'}^2/s,M_{W'}) \, B(W' \rightarrow H^+ A^0 , H^+ H^0)
\eeq
where \\[-9mm]
\beq
w(z,\mu) = \int_x^1\frac{dx}{x}\left[u(x,\mu) \bar{d}(\frac{z}{x},\mu)+\bar{u}(x,\mu) d(\frac{z}{x},\mu)\right] ~~.
\eeq
The functions $u(x,\mu)$ and $d(x,\mu)$ are the proton parton distribution functions for up- and down- quarks of the 
at factorization scale $\mu$.  Although QCD corrections to $W'$ production are usually significant
\cite{Sullivan:2002jt},  in our case they are somewhat reduced due to the smaller $\alpha_s$ at the large values of $M_{W'}$ that 
are relevant here. 
 
Figure \ref{xsec} shows the total cross section for the $p p \rightarrow W' \rightarrow H^+ A^0$ and 
$p p \rightarrow W' \rightarrow H^+ H^0$ processes at $\sqrt{s}$ = 8 and 14 TeV, with $\tan{\theta} = 0.25$. 
To compute these cross sections, we used FeynRules  \cite{Alloul:2013bka} 
for generating vertices from our Lagrangian, and input these into Madgraph 5  \cite{Alwall:2011uj} (with parton distribution functions CTEQ6L1 \cite{Pumplin:2002vw}),
which includes interference between the $W'$ and $W$ contributions.
We have set $M_{H^+} = 300$ GeV, $M_{A} = 200$ GeV; the cross sections are only weakly sensitive to  
the scalar masses as long as $W'$ is much heavier.
Figure \ref{xsec} also shows the cross sections for $p p \rightarrow Z' \rightarrow H^0 A^0$ and 
$p p \rightarrow Z' \rightarrow H^+ H^-$, for the same parameters.

\begin{figure}[t]
 \hspace{-0.2cm}\includegraphics[width=4.1in]{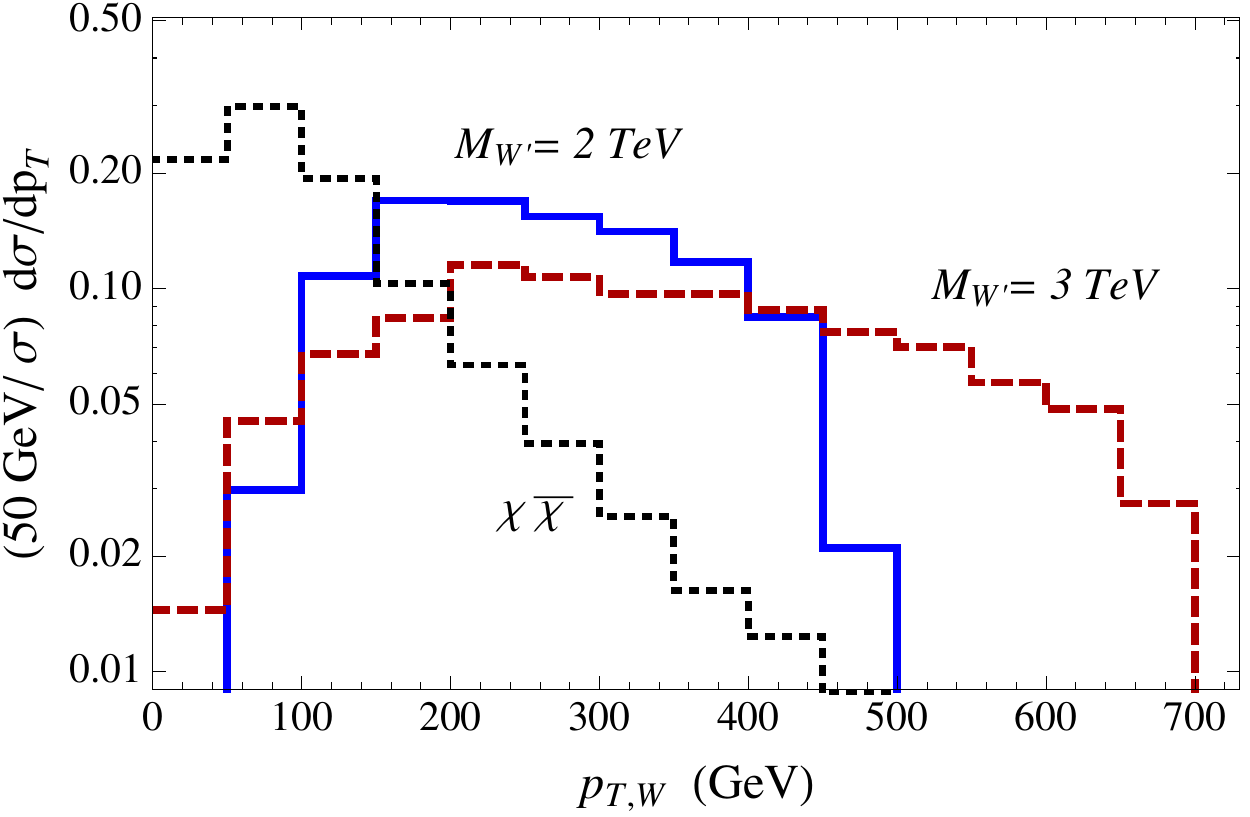} 
\caption{\label{pTW} Transverse momentum distribution of the $W$ produced 
in $p p \rightarrow W' \!\! \rightarrow H^+ A^0 \to W A^0 A^0$  when $M_{W'} = 2$ TeV (solid blue line) and 
$M_{W'} = 3$ TeV (dashed red line), for  $\sqrt{s} = 8$ TeV, $M_{H^+} = 300$ GeV and $M_{A} = 200$ GeV.
For comparison, the $W$ $p_T$ distribution (dotted black line) is included for $p p \rightarrow W \chi\bar \chi$
through a $\bar q \gamma_\mu q \, \bar \chi \gamma^\mu \chi$ 
contact interaction (for $m_\chi = 100$ GeV). }
\end{figure}

We assume that the $\mathbb{Z}_2$ symmetry discussed in Section \ref{model} is sufficiently 
preserved so that the LOP escapes the detector. As noted there, $A^0$ is most likely 
the LOP, so that each of the above processes includes two $A^0$ in the final state, which appear as 
missing transverse energy ($\,\slash\!\!\!\!E_T$) in the detector.
If $M_{W'} \gg M_{H^+}$, then the $W$ or $Z$ boson emmited in the 
cascade decays $W' \to H^+ A^0 \to W^+ A^0A^0$ and $Z' \to H^0 A^0 \to Z A^0A^0$
is highly boosted, carrying energy roughly equal of $M_{W'}/4$.
This implies that hadronic decays of the $W$ or $Z$ boson lead to 
an interesting signature with the two jets collimated into a single wide jet with substructure, plus $\,\slash\!\!\!\!E_T$.

The ATLAS collaboration \cite{Aad:2013oja} has searched for this type of signature in the case of 
DM particles pair produced through a contact interaction to quarks \cite{Beltran:2010ww,Bai:2010hh}. 
Compared to our model, the processes $pp \to W \bar\chi\chi$ and $pp \to Z \bar\chi\chi$
give rise to a smaller transverse momentum for the electroweak boson, 
which is radiated from an initial state quark.
In Figure \ref{pTW} we show the $p_T$ distributions for the $W$ 
arising from $W' \to H^+ A^0 \to W^+ A^0A^0$, as well as 
from initial state radiation
in the case of a $\bar q \gamma_\mu q \bar \chi \gamma^\mu \chi$ contact interaction (for a Dirac fermion $\chi$
of mass $m_\chi = 100$ GeV).
It is clear that the efficiency for  a stringent $p_T(W)$ cut is much higher for our $W'$ 
decays than in the case of contact interactions.

The cascade decays $W' \to H^+ H^0 \to W^+ A^0 Z A^0$ and $Z' \to H^+ H^- \to W^+A^0 W^-A^0$
lead to two highly boosted electroweak bosons plus $\,\slash\!\!\!\!E_T$. 
Hadronic decays of these $W$ and $Z$ bosons allow the use of substructure techniques to reduce the QCD 
background. 

The boosted $W$ and $Z$ ``jets" plus $\,\slash\!\!\!\!E_T$ channels have the largest branching fractions.
Nevertheless, leptonic decays of the boosted $W$ and $Z$ are also promising due to small backgrounds.
These lead to final states with one, two or three leptons, plus $\,\slash\!\!\!\!E_T$.

The mono-lepton signature has been studied theoretically \cite{Bai:2012xg} and searched for at the LHC \cite{CMS:2013iea} 
in the case of contact interactions. Again, in our case the 
$W$ producing the lepton is generically more boosted.
Unlike $W^\prime$ decays directly to a lepton-neutrino pair, there will be no Jacobian peak in the missing transverse energy distribution, as the $A^0$'s carry away a substantial fraction of the energy of the $W'$. In fact, the distribution will be peaked at low-$p_T$.  Furthermore, if the masses of the $A^0$ and $H^+$ are similar, the transverse momenta of the two final-state $A^0$ particles will have similar magnitudes but opposite directions, so their contribution to the
$\,\slash\!\!\!\!E_T$ of the event is reduced. In this case, the missing energy distribution could look like a SM $W$ decay. This problem is mitigated if the $A^0$ is substantially lighter than the Higgs triplet states, in which case the $\,\slash\!\!\!\!E_T$ distribution will have a longer tail. 

\begin{figure}[t]
\centering
\includegraphics[width=3.23in]{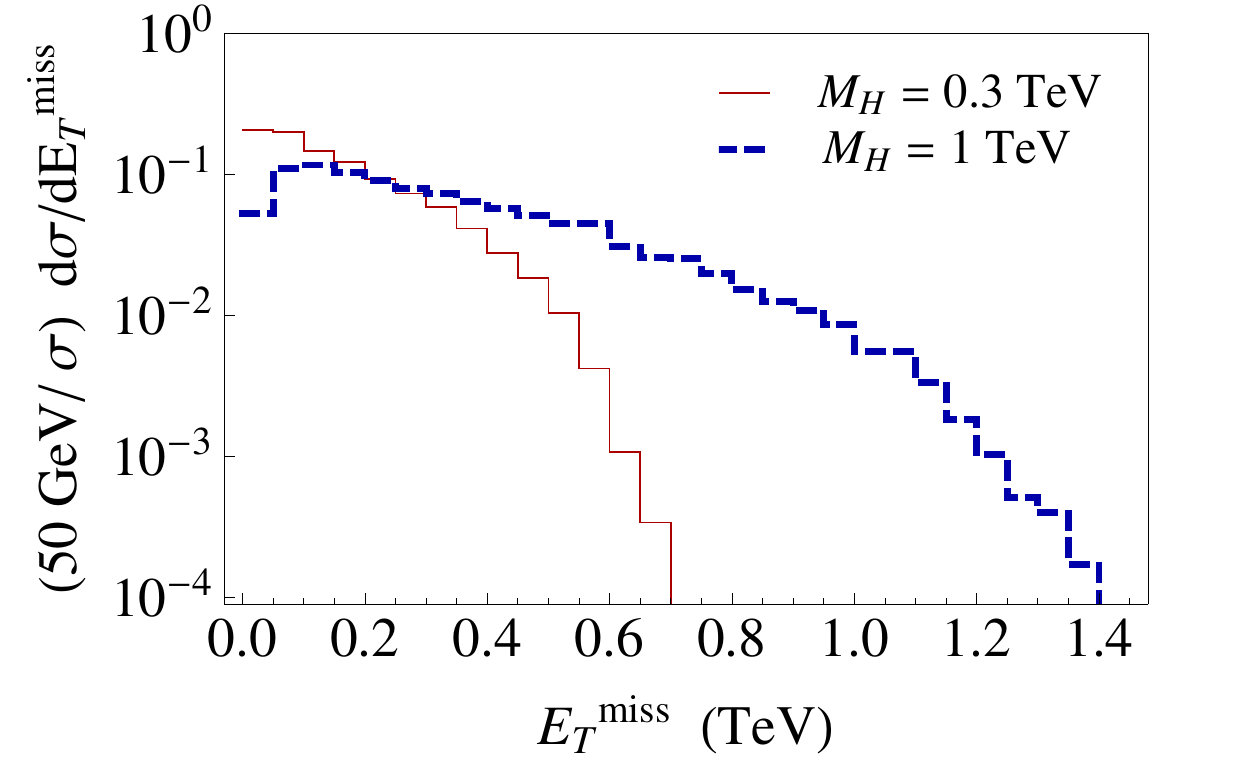} \hspace{-0.4cm}
\includegraphics[width=3.23in]{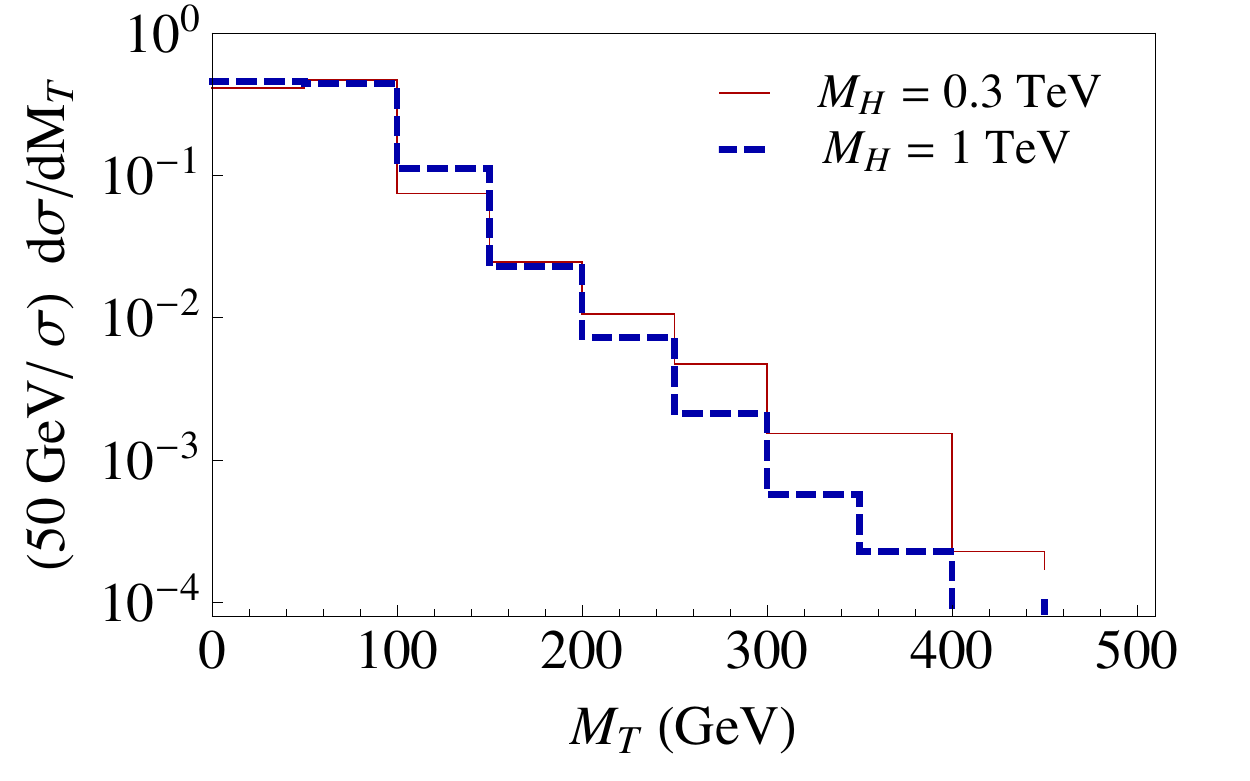} 
\caption{\label{met}  $\,\slash\!\!\!\!E_T$ distribution (left) and transverse mass (right) distributions for $M_{H^0} = 300$ GeV and $M_{H^0} = 1$ TeV, all other parameters constant ($M_{W'} = 3$ TeV, $M_{A^0} = 200$ GeV, $\tan{\theta} = 1/4$). }
\end{figure}

We  simulate $W'$ signals using Madgraph 5 \cite{Alwall:2011uj}, including showering and hadronization with 
Pythia 6.4 \cite{Sjostrand:2006za}, and PGS detector simulation \cite{PGS4}; then we analyze the events with the MadAnalysis 
package \cite{Conte:2012fm}.
Figure \ref{met} (left panel) shows missing transverse energy distributions for $M_{H^0} = 300$ GeV and $M_{H^0} = 1$ TeV, all other parameters constant. The transverse mass distribution, which is  used in LHC $W'$ searches, is also peaked at small $M_T$. Moreover, the distribution does not change substantially for different values of the Higgs masses, as shown in Figure \ref{met} (right panel). Therefore, the 
transverse mass is not the best observable for a $W'$ decaying through odd Higgs particles. 

\begin{figure}[t]
\begin{center}
\vspace{0.25cm}
\includegraphics[width=3.5in]{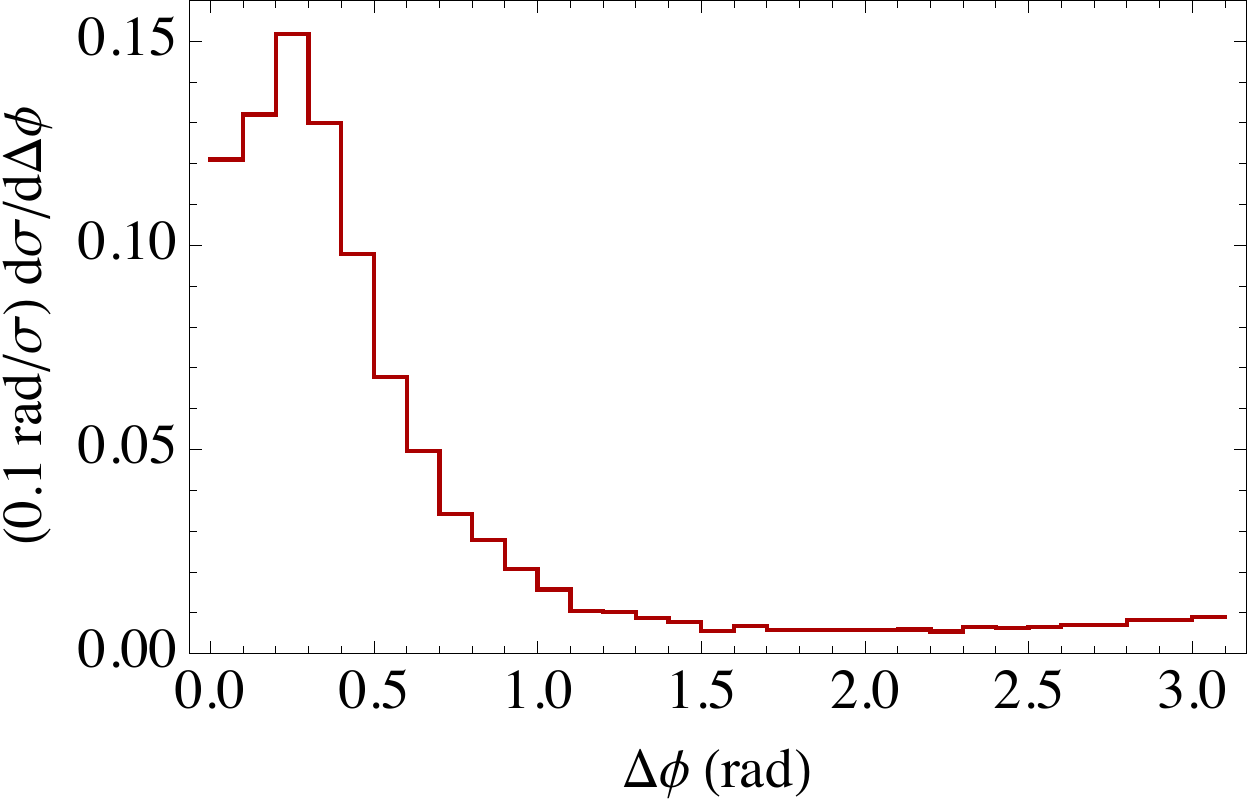}
\caption{\label{deltaphi} $\Delta \phi_{p_{T}^\ell, \,\slash\!\!\!\!E_T}$ distribution, for  $M_{H^+} = 300$ GeV, $M_{A} = 200$ GeV, and $\tan{\theta} = 0.25$.}
\end{center}
\end{figure}

A better observable for the single-lepton process $W' \rightarrow H^+A^0 \to W^+ \!+ \,\slash\!\!\!\!E_T$ is the separation in azimuthal angle between the missing transverse energy and the lepton transverse momentum, $\Delta \phi_{p_{T}^\ell, \,\slash\!\!\!\!E_T}$. When a $W$ or $W'$ decays directly to a lepton-neutrino pair, the decay products are nearly back-to-back; for both the $W'$ and dark matter mono-lepton analyses, CMS requires that $\Delta \phi_{p_{T}^\ell, \,\slash\!\!\!\!E_T} > 0.8 \pi$ \cite{CMS:2013rca}. However, the kinematics for the decay $W' \rightarrow A^0 A^0 l \nu$ are substantially different, with the $\Delta \phi_{p_{T}^\ell, \,\slash\!\!\!\!E_T}$ distribution peaked at moderate-to-small values of $\Delta \phi_{p_{T}^\ell, \,\slash\!\!\!\!E_T}$; see Figure \ref{deltaphi}. In the rest frame of the $W'$, $\vec{p}_T^{\:\l} = -\sum \vec{p}_T^{\:miss}$, but in the lab frame, the $W'$ transverse momentum is distributed among the four decay products and the correlation in azimuthal angle is lost. 

There are also two processes leading to $\ell^+\ell^-+\,\slash\!\!\!\!E_T$.
One of them is the $Z' \to H^0 A^0 \to Z A^0 A^0$ cascade decay, with $Z\to \ell^+\ell^-$; the related process in the case of contact interactions has been discussed in \cite{Carpenter:2012rg}. 
The other one is $Z'\! \to H^+ H^-\! \to W^+A^0 \, W^-A^0$ with leptonic $W$ decays; a similar final state, but without $s$-channel resonance,
arises from chargino pair production Ref.~\cite{Curtin:2012nn}.

\begin{figure}[t]
\begin{center}
\includegraphics[width=3.5in]{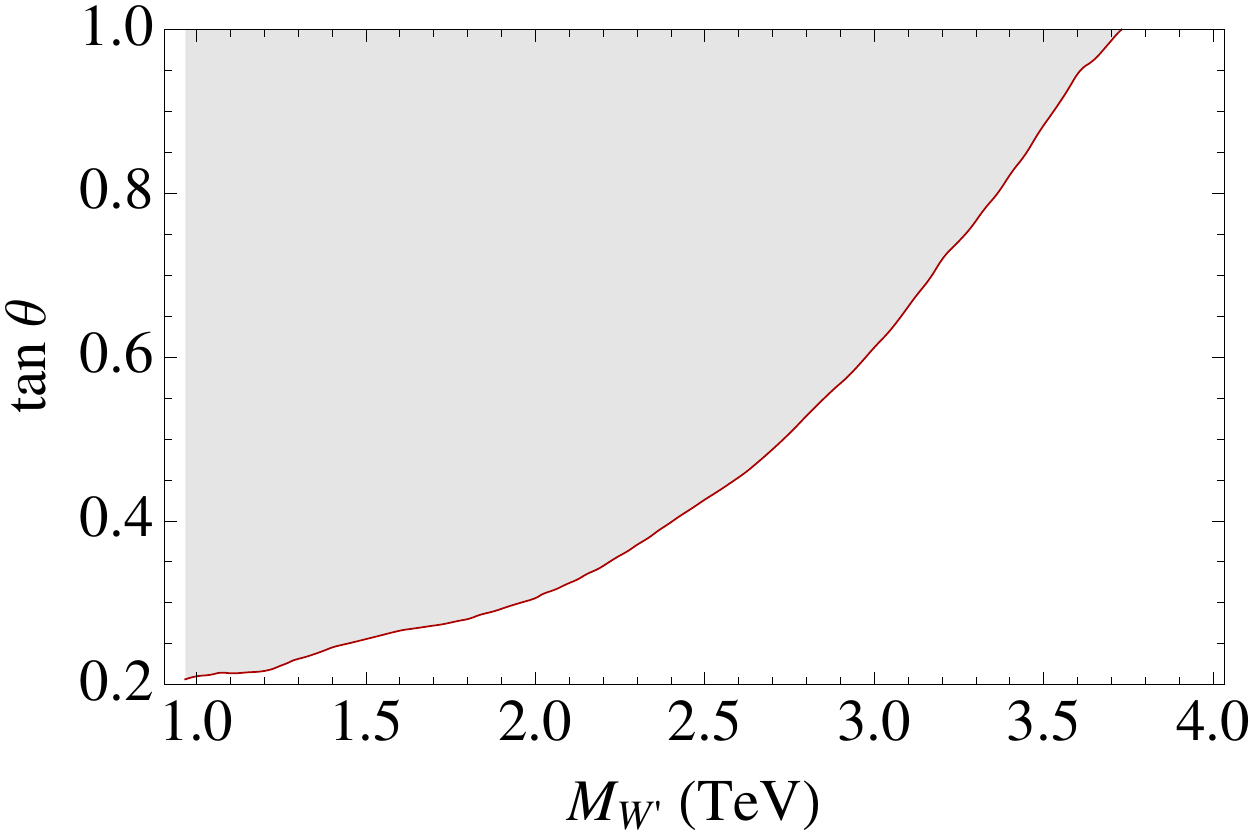}
\caption{\label{limits} Exclusion limit in the $M_{W'}-\tan{\theta}$ plane, derived from the CMS 
$W' \rightarrow \ell \nu$ search \cite{CMS:2013rca}, for $M_{H} = 300$ GeV and $M_{A} = 200$ GeV.}
\end{center}
\end{figure}

The  limits on our model  set by current LHC results are already stronger than those from electroweak fits mentioned in
Section \ref{model}. The searches in the $W' \rightarrow \ell \nu$ channel, although 
affected by suppressed branching fraction for small $\tan{\theta}$,
set relevant bounds.  In Figure \ref{limits}, we 
reinterpret the 95\% CL limit set by the CMS Collaboration \cite{CMS:2013rca}
on $\sigma_{excl.}/\sigma_{{\rm SSM} W'}$ as a limit on $\tan{\theta}$.  

Existing LHC searches for other processes set less stringent limits.
For $W' \rightarrow H^+ A^0 \to W^+ A^0A^0 \to \ell + \,\slash\!\!\!\!E_T$, we use Figure 4 of \cite{CMS:2013iea} to estimate the number of background events with $1 < \Delta \phi_{p_{T}^\ell, \,\slash\!\!\!\!E_T} < 1.5$ in the muon channel, then set an upper limit on $W' \rightarrow \mu + \,\slash\!\!\!\!E_T$ events in the same region assuming no excess is observed. 
This limit is at most $M_{W'} > 1.05$ TeV for any $\tan\theta \leq 1$. 

For the process $W' \rightarrow H^+ H^0 \to W^+ A^0 Z A^0 \to \ell^+\ell^+\ell^- + \,\slash\!\!\!\!E_T$, we use results from leptonic searches for charginos and neutralinos in \cite{CMS:2013dea}. We consider the search for a same-flavor, opposite sign electron or muon pair on the $Z$ peak ($75 \text{ GeV} < M_{\ell\ell} < 105 \text{ GeV}$), plus an additional electron or muon.  We sum over $M_T$ bins, then we use a Poisson likelihood function multiplied over $\,\slash\!\!\!\!E_T$ bins to set an upper bound on the number of events. The upper limits on the number of $W'$ events in each channel are then translated to a limit on $\tan{\theta}$ as a function of $M_{W'}$. This limit is rather weak: for $M_{W'} =1$ TeV,
only values of  $\tan{\theta} > 3.5$ are excluded.

We see that the search for direct decays to lepton plus $\,\slash\!\!\!\!E_T$ final states still provides the most stringent constraint on our model. The reason is that the 3-lepton rate is suppressed by both the $W\rightarrow \ell \nu$ and $Z \rightarrow \ell^+\ell^-$ branching fractions.  
Furthermore, for the mono-lepton search, the selection cuts that optimize signal over background for $W\rightarrow \ell \nu$ cut out a substantial portion of the $H^\pm A^0$ events.  A new analysis focusing on the small 
$ \Delta \phi_{p_{T}^\ell, \,\slash\!\!\!\!E_T}$ region, using both the electron and muon channels, would provide a stronger limit. 

Given the mass degeneracy between $W'$ and $Z'$, 
limits set by searches for $Z^\prime \to \ell^+\ell^-$ can also be plotted in the $M_{W'}-\tan{\theta}$ plane.
However, they are weaker than those from $W' \rightarrow \ell \nu$ because both the production 
cross section and the leptonic branching fraction are smaller for $Z'$ than for $W'$.

\setcounter{equation}{0}
\section{\label{conclusions} Conclusions}

The $SU(2)_1 \times SU(2)_2 \times U(1)_Y$ model with a bidoublet
and a doublet complex scalars is a simple renormalizable model that can serve as 
a benchmark for various LHC searches.
It includes a meta-sequential $W'$ boson whose $s$-channel production interferes constructively with 
the $W$ contribution, and depends on only two parameters: $M_{W'}$ and the 
overall coupling normalization, $\tan\theta$. 
It also includes a $Z^\prime$ boson (degenerate in mass with $W'$) 
which couples, to a good approximation, only to left-handed fermions.

The potential for the bidoublet ($\Delta$) and doublet ($\Phi$) scalars
is chosen to be invariant under a $\mathbb{Z}_2$ transformation
that interchanges the bidoublet and its charge conjugate.
The physical scalar spectrum then consists of four odd Higgs particles
(a mass-degenerate weak-triplet $H^+,H^0,H^-$, and a CP-odd singlet $A^0$),
the recently discovered Higgs boson ($h^0$), and a heavier scalar ($H^\prime$) whose 
couplings to SM fields are the same as those of $h^0$ except for an universal suppression.
The $A^0$ is naturally the LOP because a global $U(1)$ symmetry becomes exact 
in the $M_A \to 0$ limit.

The phenomenology of the scalars is worth exploring whether or not the $W'$ and $Z'$
bosons are light enough to be produced at the LHC. Electroweak production of the triplet scalars,
for example, would lead to final states involving one or two weak bosons and two LOPs.

The range of parameters where $A^0$ is a viable dark matter particle remains to be studied.
For the present work we focused on the case where $A^0$ is sufficiently long-lived to escape 
the detector, but we also mentioned possible signatures in the case where $A^0$ decays 
(promptly or with a displaced vertex) into fermion pairs. 

This model illustrates nicely the possibility that the $W'$ and $Z'$ bosons may decay predominantly (with branching fraction as large as 96\%) into
the scalars responsible for breaking the extended gauge symmetry.
Generically, the high-energy behavior of any $W'$ boson 
requires it to be associated with a non-Abelian gauge symmetry (or else it must be a bound state with 
the compositeness scale not much higher than its mass), which in turn implies a larger Higgs sector.
The non-Abelian gauge coupling can be significantly larger than the Higgs quartic couplings,
implying vector bosons much heavier than the scalars. 

In our model, the $W'$ and $Z'$ couplings to the odd Higgs particles are enhanced for $\tan\theta \ll 1$ by $1/\tan\theta$. Consequently, the usual $u\bar{d} \to W' \rightarrow \ell \nu$ or $t\bar b$
channels currently used in searches at the LHC are suppressed both in production and in 
braching fractions, the combined effect being of order $\tan^6\!\theta$. The mass limits on a sequential 
$W'$, currently around 3.8 TeV, are relaxed for $\tan\theta \approx 0.2$ (the lower perturbativity bound)
to $M_{W'} > 1$ TeV.
At the same time, the cascade decays through odd Higgs particles, 
$W'\! \rightarrow\! H^+ A^0 \!\to\! W^+\! A^0 A^0$, \ 
$W'\! \rightarrow\! H^+ H^0 \!\to\! W^+\! A^0 \,Z A^0$, \
$Z' \! \!\rightarrow \!H^0 A^0 \!\to\! Z A^0 A^0$ and 
$Z' \! \!\rightarrow\! H^+ H^- \! \!\to\! W^+\! A^0 \, W^- A^0$
allow interesting searches at the LHC, with boosted $W$ and $Z$ bosons decaying either hadronically or leptonically. \\[4mm]

{\bf Acknowledgments:} \
We would like to thank Calin Alexa, Yang Bai, Vernon Barger, Patrick Fox, Adam Jinaru, Sudhir Malik, Caroline Milstene and Liantao Wang for stimulating conversations.
A.P. is supported by the Fermilab Fellowship in Theoretical Physics. Fermilab is operated by Fermi Research Alliance, LLC, under Contract No.~DE-AC02-07CH11359 with the US Department of Energy. 



\begin{thebibliography}{99} \frenchspacing

\bibitem{Beringer:1900zz} 
  J.~Beringer {\it et al.}  [Particle Data Group Collaboration],
  ``Review of Particle Physics (RPP),''
  Phys.\ Rev.\ D {\bf 86}, 010001 (2012).

\bibitem{Mohapatra:1986uf} 
See {\it e.g}, R.~N.~Mohapatra,
  ``Unification and supersymmetry. The frontiers of quark - lepton physics,''
  New York, USA: Springer (2003) 421 p;
  J.~L.~Hewett and T.~G.~Rizzo,
  ``Low-energy phenomenology of superstring inspired $E_6$ models,''
  Phys.\ Rept.\  {\bf 183}, 193 (1989);
  P.~Langacker and S.~U.~Sankar,
  ``Bounds on the mass of $W_R$ and the $W_L-W_R$ mixing angle $\xi$ in general $SU(2)_L \times SU(2)_R \times U(1)$ models,''
  Phys.\ Rev.\ D {\bf 40}, 1569 (1989);
  V.~D.~Barger, K.~Whisnant and W.-Y.~Keung,
  ``Gauge model with small $C_q^2$ interaction term and heavy weak bosons,''
  Phys.\ Rev.\ D {\bf 25}, 291 (1982);
  K.~Hsieh, K.~Schmitz, J.-H.~Yu and C.-P.~Yuan,
  ``Global analysis of general $SU(2) \times SU(2) \times U(1)$ models with precision data'',
  Phys.\ Rev.\ D {\bf 82}, 035011 (2010)
  [arXiv:1003.3482];
  T.~Abe, N.~Chen and H.~-J.~He,
  ``LHC Higgs signatures from extended electroweak gauge symmetry'',
  JHEP {\bf 1301}, 082 (2013)
  [arXiv:1207.4103].
   
\bibitem{CMS:2013rca} 
  CMS Collaboration,
  ``Search for leptonic decays of $W'$ bosons in $pp$ collisions at $\sqrt{s}=8$ TeV'', report
  CMS-PAS-EXO-12-060, March 2013.

\bibitem{Barger:1980ix} 
  V.~D.~Barger, W.-Y.~Keung and E.~Ma,
  ``A gauge model with light $W$ and $Z$ bosons,''
  Phys.\ Rev.\ D {\bf 22}, 727 (1980);
  M.~Schmaltz and C.~Spethmann,
  ``Two simple $W'$ models for the early LHC,''
  JHEP {\bf 1107}, 046 (2011)
  [arXiv:1011.5918].
  
\bibitem{Cao:2012ng} 
  Q.-H.~Cao, Z.~Li, J.-H.~Yu and C.~P.~Yuan,
  ``Discovery and identification of $W'$ and $Z'$ in $SU(2) \times SU(2) \times U(1)$  models at the LHC,''
  Phys.\ Rev.\ D {\bf 86}, 095010 (2012)
  [arXiv:1205.3769].

\bibitem{Dodelson} 
  S.~Dodelson, B.~R.~Greene and L.~M.~Widrow,
  ``Baryogenesis, dark matter and the width of the $Z$'',
  Nucl.\ Phys.\ B {\bf 372}, 467 (1992);
  G.~Belanger, B.~Dumont, U.~Ellwanger, J.~F.~Gunion and S.~Kraml,
  ``Global fit to Higgs signal strengths and couplings and implications for extended Higgs sectors,''
  Phys.\ Rev.\ D {\bf 88}, 075008 (2013)
  [arXiv:1306.2941].

\bibitem{Cirelli:2005uq} 
  M.~Cirelli, N.~Fornengo and A.~Strumia,
  ``Minimal dark matter,''
  Nucl.\ Phys.\ B {\bf 753}, 178 (2006)
  [hep-ph/0512090].

\bibitem{Barbieri:2006dq} 
  R.~Barbieri, L.~J.~Hall and V.~S.~Rychkov,
  ``Improved naturalness with a heavy Higgs: An alternative road to LHC physics,''
  Phys.\ Rev.\ D {\bf 74}, 015007 (2006)
  [hep-ph/0603188].

\bibitem{poster}
A.~Jinaru, C.~Alexa, I.~Caprini and A.~Tudorache, poster presented at 
the International Conference on New Frontiers in
Physics (Kolymbari, Creta), Sept. 2013.

\bibitem{Sullivan:2002jt} 
  Z.~Sullivan,
  ``Fully differential $W^\prime$ production and decay at next-to-leading order in QCD,''
  Phys.\ Rev.\ D {\bf 66}, 075011 (2002)
  [hep-ph/0207290].
  
\bibitem{Alloul:2013bka} 
  A.~Alloul, N.~D.~Christensen, C.~Degrande, C.~Duhr and B.~Fuks,
  ``FeynRules 2.0 - A complete toolbox for tree-level phenomenology,''
  arXiv:1310.1921 [hep-ph]; 
  C.~Degrande, C.~Duhr, B.~Fuks, D.~Grellscheid, O.~Mattelaer and T.~Reiter,
  ``UFO - The Universal FeynRules Output,''
  Comput.\ Phys.\ Commun.\  {\bf 183}, 1201 (2012)
  [arXiv:1108.2040 [hep-ph]].

\bibitem{Alwall:2011uj} 
  J.~Alwall, M.~Herquet, F.~Maltoni, O.~Mattelaer and T.~Stelzer,
  ``MadGraph 5 : Going Beyond,''
  JHEP {\bf 1106}, 128 (2011)
  [arXiv:1106.0522 [hep-ph]].

\bibitem{Pumplin:2002vw} 
 J.~Pumplin, D.~R.~Stump, J.~Huston, H.~L.~Lai, P.~M.~Nadolsky and W.~K.~Tung,
  ``New generation of parton distributions with uncertainties from global QCD analysis,''
  JHEP {\bf 0207}, 012 (2002)
  [hep-ph/0201195].

\bibitem{Aad:2013oja} 
  G.~Aad {\it et al.}  [ATLAS Collaboration],
  ``Search for dark matter in events with a hadronically decaying $W$ or $Z$ 
  boson and missing transverse momentum in $pp$ collisions at $\sqrt{s}$=8 TeV'',
  arXiv:1309.4017 [hep-ex].
  
\bibitem{Beltran:2010ww} 
  M.~Beltran, D.~Hooper, E.~W.~Kolb, Z.~A.~C.~Krusberg and T.~M.~P.~Tait,
  ``Maverick dark matter at colliders,''
  JHEP {\bf 1009}, 037 (2010)
  [arXiv:1002.4137 [hep-ph]].
  
\bibitem{Bai:2010hh} 
  Y.~Bai, P.~J.~Fox and R.~Harnik,
  ``The Tevatron at the frontier of dark matter direct detection,''
  JHEP {\bf 1012}, 048 (2010)
  [arXiv:1005.3797];
 
\bibitem{Bai:2012xg} 
  Y.~Bai and T.~M.~P.~Tait,
  ``Searches with mono-leptons,''
  Phys.\ Lett.\ B {\bf 723}, 384 (2013)
  [arXiv:1208.4361].
  
\bibitem{CMS:2013iea} 
  CMS Collaboration,
  ``Search for dark matter in the mono-lepton channel with 
$pp$ collision events at center-of-mass energy of 8 TeV'', report 
  CMS-PAS-EXO-13-004, July 2013.

 
\bibitem{Sjostrand:2006za} 
  T.~Sjostrand, S.~Mrenna and P.~Z.~Skands,
  ``PYTHIA 6.4 physics and manual,''
  JHEP {\bf 0605}, 026 (2006)
  [hep-ph/0603175].

\bibitem{PGS4}
  J.~Conway, 
   R.~Culbertson, R.~Demina, B.~Kilminster, M.~Kruse, S.~Mrenna, J.~Nielsen, M.~Roco,
  ``Pretty Good Simulation of high energy collisions",\\
   {\tt http://physics.ucdavis.edu/}$\small \sim${\tt conway/research/software/pgs/pgs4-general.htm}

\bibitem{Conte:2012fm} 
  E.~Conte, B.~Fuks and G.~Serret,
  ``MadAnalysis 5, a user-friendly framework for collider phenomenology,''
  Comput.\ Phys.\ Commun.\  {\bf 184}, 222 (2013)
  [arXiv:1206.1599 [hep-ph]].
  
\bibitem{CMS:2013dea} 
  CMS Collaboration,
  ``Search for electroweak production of charginos, neutralinos, and sleptons using leptonic final states in $pp$ collisions at 8 TeV'', report 
  CMS-PAS-SUS-13-006, July 2013; \ 
  ATLAS Collaboration,
  ``Search for direct production of charginos and neutralinos in events with three leptons and missing transverse momentum in 21$\,$fb$^{-1}$ of pp collisions at $\sqrt{s}=8$ TeV'', report
  ATLAS-CONF-2013-035, April 2013.

\bibitem{Carpenter:2012rg} 
  L.~M.~Carpenter, A.~Nelson, C.~Shimmin, T.~M.~P.~Tait and D.~Whiteson,
  ``Collider searches for dark matter in events with a $Z$ boson and missing energy,''
  arXiv:1212.3352.

\bibitem{Curtin:2012nn}
  G.~D.~Kribs, A.~Martin and T.~S.~Roy,
  ``Supersymmetry with a chargino NLSP and gravitino LSP,''
  JHEP {\bf 0901}, 023 (2009)
  [arXiv:0807.4936];
  D.~Curtin, P.~Jaiswal and P.~Meade,
  ``Charginos hiding in plain sight,''
  Phys.\ Rev.\ D {\bf 87}, no. 3, 031701 (2013)
  [arXiv:1206.6888].

\end{thebibliography}
\end{document}